\documentclass[aps,pra,preprint,superscriptaddress]{revtex4-1}
\usepackage{amsmath,amssymb,amsfonts}
\usepackage[english]{babel}
\usepackage{graphicx}
\usepackage{dcolumn}
\usepackage{bm}
\usepackage{color}

\begin{document}

\title{Quantum phase transition of nonlinear light in the finite size Dicke Hamiltonian}

\author{B.~M. Rodr\'{\i}guez-Lara}
\email[]{bmlara@mx.nthu.edu.tw}
\affiliation{Institute of Photonic Technologies, National Tsing-Hua University, Hsinchu 300, Taiwan.}
\author{Ray-Kuang Lee}
\email[]{rklee@mx.nthu.edu.tw}
\affiliation{Institute of Photonic Technologies, National Tsing-Hua University, Hsinchu 300, Taiwan.}

\date{\today}

\begin{abstract}
We study the quantum phase transition of a $N$ two-level system ensemble interacting with an optical degenerate parametric process, which can be described by the finite size Dicke Hamiltonian plus counter-rotating and quadratic field terms. 
Analytical closed forms of the critical coupling value and their corresponding separable ground states are derived in the weak and strong coupling regimes. 
The existence of bipartite entanglement between the two-level-system ensemble and photon field as well as between ensemble components for moderate coupling is shown through numerical analysis. 
Given a finite size, our results also indicate the co-existence of squeezed fields and squeezed atomic ensembles.
\end{abstract}

\maketitle
\section{Introduction}

The study of light--matter interaction has been the central topic of quantum optics; it has laid the foundation for laser theory, quantum state engineering, fundamental testing of quantum mechanics, and implementation of quantum information processing \cite{Cohen2004}.
Among the various systems involving the interaction of photons and atoms, the simplest and the most important building block to illustrate interesting quantum phenomena involves just the one two-level atom (TLA) \cite{Allen1975}. As the number of TLAs increases, collective effects give rise to intriguing many-body phenomena; \textit{e.~g.} the existence of a coherent super-radiant phase at zero temperature \cite{Dicke1954p99}.

The Hamiltonian describing the coupling of a non-interacting atomic ensemble with a single quantized electromagnetic field mode is equivalent, via a Power-Zineau transformation, to the Hamiltonian of a free particle under a field related potential \cite{Cohen2004}. The ground state energy of such a system is bounded from below by the atomic ground state and the vacuum field, \textit{ergo} a super-radiant phase transition of the ground state in a charge only system coupled to a single field mode is not possible  \cite{BialynickiBirula1979p301,Rzazewski1991p593}. Within the standard minimal coupling, long wavelength and rotating wave approximation, and discarding quadratic terms, the interaction of a photon field with an ensemble of TLAs is described by the Dicke Hamiltonian,
\begin{equation} \label{eq:DickeH}
\hat{H}_{\text{Dicke}} = \hbar \omega_{p} \hat{a}^{\dagger} \hat{a} +  \hbar \omega_{a} \hat{J}_{z} + \hbar \frac{\lambda}{\sqrt{N}} ( \hat{a} \hat{J}_{+} + \hat{a}^{\dagger} \hat{J}_{-}),
\end{equation}
where the transition energy for each one of the $N$ TLAs and the radiation field frequency are $\omega_a$ and $\omega_p$, in that order. The atomic ensemble operators are defined in terms of the Pauli matrices for the $j$-th TLA as $\hat{J}_{z} = (1/2) \sum_{j=1}^{N} \sigma_{z}^{(j)}$ and $\hat{J}_{\pm} = \sum_{j=1}^{N} \sigma_{\pm}^{(j)}$. The coupling strength for the photon-atom interaction is denoted by $\lambda$.
The impossibility for a quantum phase transition (QPT) in the source model translates into a violation of Thomas-Reiche-Kuhn (TRK) sum rule for an atom if the relation between coupling strength and quadratic parameter for a phase transition in the Dicke model were to occur \cite{Rzazewski1975p432}. In order to observe the effect of radiation-matter coupling known as super-radiant phase transition in charge only systems, more realistic models, such as spin magnetic moment, statistics or infinitely many electromagnetic field modes, should be taken into account  \cite{Gawedzki1981p2134}.

Despite the fact that a QPT is forbidden in the physical system that originated Dicke model, the phenomenon is interesting by itself. First, the existence of a QPT in the Dicke Hamiltonian was reported as a series of instabilities of the ground state for a finite size ensemble of two-level systems \cite{Mallory1969p1976}. Then, the existence of a super-radiant thermodynamic phase transition was proved for an infinitely large ensemble interacting with a coherent boson field at a given temperature \cite{Hepp1973p360,Wang1973p831,Hepp1973p2517}. Also, in the classical limit, studies indicated that this quantum critical phenomenon is associated to quantum chaos and ensemble-field entanglement \cite{Emary2003p044101,Emary2004p053804,Lambert2004p073602}. Non-trivial scaling exponents at the critical point have been discussed on the large ensemble size regime \cite{Vidal2006p817}. Bipartite intra-ensemble entanglement, due to finite size effects, was demonstrated for the Dicke Hamiltonian \cite{Buzek2005p163601}. 
This theoretical understanding has motivated proposals for the realization of Dicke model in systems that might allow a super-radiant phase transition; \textit{e.g.} open dynamical systems involving semiconductor quantum wells or quantum dots \cite{Brandes1998p3952, Vorrath2003p035309}, open dynamical cavity-QED systems with neutral atoms \cite{Dimer2007p013804} and ions \cite{Harkonen2009p033841}, and superconducting quantum devices \cite{Blais2004p062320,Chen2007p055803}, just to mention a few. Recently, it has been shown that a standing-wave laser driven Bose-Einstein condensate (BEC) coupled to a high finesse optical cavity, accounting for the center of mass motion, realizes the Dicke model and that the superradiant phase corresponds to a periodical self-organized phase of the atoms \cite{Baumann2010p1301,Nagy2010p130401}.

Currently, there is great interest in pursuing quantum phase transitions of light since photons interacting with atoms should be much easier to study and probe than electrons in condensed matter systems \cite{Illuminati2006p803,Greentree2006p856}. Schemes to realize composite Dicke models have been proposed; \textit{e.g.} by combining photon hopping between identical cavities in the photon-blockade regime, Mott-insulator to superfluid QPT has been demonstrated in the Dicke-Bose-Hubbard model for an arbitrary number of two-level atoms \cite{Lei2008p033827}. More exotic QPTs of light have been predicted in a Heisenberg spin $1/2$ Hamiltonian \cite{Hartmann2007p160501}, two species Bose-Hubbard model \cite{Hartmann2008p033011}, arrays of coupled cavities \cite{Rossini2007p186401,Angelakis2007p031805}, and dual-species optical-lattice cavity \cite{Lei2010}. These studies have brought the possibility to analyze critical quantum phenomena in conventional condensed matter systems by manipulating the interaction between photons and atoms.

As nonlinear optics plays an important role in quantum optics, especially in the generation of quantum noise squeezed states \cite{Scully1997, Yamamoto1999}, a natural question one may ask is how to associate nonlinear quantum processes with the phenomenon of QPT. Sub-Poissonian photon statistics of the field state, and momentum squeezing of the atomic state, have been predicted for the Dicke model \cite{Lambert2004p073602}. It also has been proposed that squeezing of the photon field carries signatures of the associated quantum critical phenomena in the size-consistent Dicke model \cite{Jarrett2007p34001}.

In this work, we study the quantum critical phenomena of $N$ two-level systems embedded within a nonlinear optical medium, with $N$ finite. An optical degenerate parametric down conversion (PDC) process, where the nonlinear medium is pumped by a classical field of frequency $2 \omega_f$ and that field is converted into pairs of identical photons of frequency $\omega_f$ each, is considered. The corresponding nonlinear interaction Hamiltonian is given by 
\begin{eqnarray} \label{eq:PDC}
\hat{H}_{\text{PDC}}&=& \hbar \kappa (\hat{a}^2+ \hat{a}^{\dagger 2}),
\end{eqnarray}
where the nonlinear parameter $\kappa = \chi^{(2)}\beta$ is defined by the second-order nonlinearity coefficient $\chi^{(2)}$ and the classical amplitude of the pumping field $\beta$. By plugging the degenerate PDC Hamiltonian, Eq.(\ref{eq:PDC}), into the Dicke model, Eq.(\ref{eq:DickeH}), and restoring the counter-rotating terms we obtain the following Hamiltonian,
\begin{eqnarray}
\label{eq:DmH}
\hat{H}&=& \hbar \omega_{f} \hat{a}^{\dagger} \hat{a} +  \hbar \omega_{a} \hat{J}_{z} + \hbar \frac{\lambda}{\sqrt{N}} ( \hat{a} + \hat{a}^{\dagger})\hat{J}_{x}   + \hbar \kappa (\hat{a} + \hat{a}^{\dagger})^2,
\end{eqnarray}
where we have rescaled the system energy by $(-\hbar \kappa)$ and redefined the photon frequency as $\omega_f = \omega_p - 2\,\kappa$. In the literature, the Hamiltonian in Eq. (\ref{eq:DmH}) is that of a non-interacting TLAs ensemble driven by an electromagnetic field with the standard minimal coupling in the long wavelength limit. Of course, such a QPT does not exist for charge-only systems interacting with a finite number of electromagnetic radiation modes \cite{Rzazewski1975p432,BialynickiBirula1979p301,Gawedzki1981p2134,Rzazewski1991p593} but the proposed nonlinear optics PDC process may provide a strong nonlinearity for the field, which might be coupled to a feasible realization of Dicke model \cite{Brandes1998p3952, Vorrath2003p035309,Dimer2007p013804,Harkonen2009p033841,Blais2004p062320,Chen2007p055803,Baumann2010p1301,Nagy2010p130401,Knight1978p1454,Lee2004p083001,Keeling2007p295213}.
The theoretical quantum phase transition of the proposed model, Eq. (\ref{eq:DmH}), has been shown in the thermodynamic limit, \textit{i.~e.} both ensemble size and volume are considered infinitely large, $N$ and $V \rightarrow \infty$, \cite{Hioe1973p1440}.  To the knowledge of the authors, the finite size effect on the QPT of this Hamiltonian remains unanswered.

The purpose of this work is twofold. First, we show analytically that a pair of unitary transformations effect a rotating wave approximation equivalence on the Hamiltonian in the weak coupling regime, thus the exact critical coupling strength can be calculated \cite{Buzek2005p163601}. In the strong coupling regime, we show how the critical coupling strength can be calculated under just a semi-classical field approximation. In both cases, the critical coupling value of the finite-size $N$ Dicke Hamiltonian plus counter-rotating and quadratic field terms is independent of the atomic number and agrees with results derived from the classical limit, $N$ and $V \rightarrow \infty$.
Second, treated as a degenerate parametric process, we numerically demonstrate the existence of squeezed fields and squeezed atomic ensembles as well as bipartite entanglement between the two-level system ensemble and photon field and among ensemble components themselves.

\section{Weak coupling regime, $\lambda \ll \omega_{a}$.}
In order to derive a set of critical coupling values for the quantum phase transition, we first consider the weak coupling regime, \textit{i.~e.}, $\lambda \ll \omega_{a}$. In analogy to the unitary squeezed operator for a degenerate PDC, we define the following unitary transformation and associated parameter,
\begin{eqnarray}
\hat{T} = e^{\eta \left( \hat{a}^{2} - \hat{a}^{\dagger 2} \right)}, \quad \eta = \kappa / \left[2 (\omega_{f} + 2 \kappa) \right].
\end{eqnarray}
Under the restriction $\eta \ll 1$, the Hamiltonian in Eq. (\ref{eq:DmH}) is reduced into 
\begin{eqnarray} \label{eq:WH1}
\tilde{H} &=& \hat{T}^{-1}  \hat{H} \hat{T}, \nonumber  \\
&\approx& \hbar \tilde{\omega}_{f} \hat{a}^{\dagger} \hat{a} +  \hbar \frac{\omega_{a}}{2} \hat{J}_{z} + \hbar \frac{\tilde{g}}{\sqrt{N}} (\hat{a} + \hat{a}^{\dagger}) \hat{J}_{x} + \hbar \tilde{\kappa}.
\end{eqnarray}
Thus, the original Hamiltonian, Eq. (\ref{eq:DmH}), is approximated by the well-known Dicke Hamiltonian plus the counter rotating terms with a modified field frequency $\tilde{\omega}_{f}$, a modified coupling constant $\tilde{g}$, and a constant energy shift $\hbar \tilde{\kappa}$, defined as
\begin{eqnarray}
\tilde{\omega}_{f} = \left(\frac{\omega_{f}+ 4 \kappa}{\omega_{f}+ 2 \kappa}\right)\omega_{f}, \quad 
\tilde{g} = \left( \frac{\omega_{f} + \kappa}{\omega_{f}+ 2 \kappa} \right) \lambda,  \quad
\tilde{\kappa} = \left( \frac{\omega_{f}}{\omega_{f} + 2 \kappa} \right)\kappa.
\end{eqnarray}
As the total excitation number, $\hat{N} = \hat{a}^{\dagger}\hat{a} + \hat{J}_{z}$, does not commute with this Hamiltonian, Eq.\eqref{eq:WH1}, in order to further simplify our problem, a second unitary transformation is used,
\begin{equation}
\hat{U} = e^{- \imath \xi ( \hat{a} + \hat{a}^{\dagger}) \hat{J}_{y}}, \quad \xi = \tilde{g}/[\sqrt{N} (\omega_{a} + \tilde{\omega}_{f})],
\end{equation}
where the newly defined parameter fulfills $\xi \ll 1$ due to the weak coupling regime requirement $\lambda \ll \omega_{a}$. Neglecting all but linear powers of the parameter $\xi$, the Hamiltonian in Eq. (\ref{eq:DmH}) is written
\begin{eqnarray}
\hat{H}_{W} &=& \hat{U}^{-1} \hat{T}^{-1} \hat{H} \hat{T} \hat{U}, \nonumber \\
&\approx& \hbar \tilde{\omega}_{f}  \hat{a}^{\dagger} \hat{a} +  \hbar \left[ \omega_{a} + \tilde{\omega}_{a} (\hat{a} + \hat{a}^{\dagger})^2 \right] \hat{J}_{z} + \hbar\frac{\tilde{\lambda}}{\sqrt{N}} \left( \hat{a} \hat{J}_{+} + \hat{a}^{\dagger} \hat{J}_{-} \right), \label{eq:HamWeakLimit}
\end{eqnarray}
with the extra frequency $\tilde{\omega_{a}}$ and the modified coupling $\tilde{\lambda}$ given by the expressions
\begin{eqnarray}
\tilde{\omega}_{a} = \frac{ 2 \tilde{g}^2}{N(\omega_{a} + \tilde{\omega}_{f})} , \quad
\tilde{\lambda} = \frac{2 \tilde{\omega}_{f} \tilde{g}}{(\omega_{a} + \tilde{\omega}_{f})}.
\end{eqnarray}
The weak regime assumption makes it possible to neglect the extra frequency $\tilde{\omega}_{a}$. Thus the weak limit Hamiltonian, Eq.\eqref{eq:HamWeakLimit}, is further reduced to the well-known finite size Dicke model in Eq. (\ref{eq:DickeH}), of which the ground state can be found exactly and undergoes a phase transition at the critical value $ \tilde{\lambda} = \sqrt{\omega_{a}\tilde{\omega}_f}$ \cite{Buzek2005p163601, Tsyplyatyev2009p012134}.
In our case, the critical coupling  value in the weak coupling regime can be explicitly expressed as
\begin{equation} \label{eq:WCritC}
\lambda_{W} \approx \frac{\omega_{f} (\omega_{a} + \omega_{f}) + 2 \kappa (\omega_{a} + 2 \omega_{f})}{2 (\omega_{f} + \kappa)} \sqrt{ \frac{\omega_{a} (\omega_{f} + 2 \kappa)}{\omega_{f} ( \omega_{f} + 4 \kappa )}}. 
\end{equation}
For coupling values fulfilling the condition $\lambda \ll \lambda_{W}$, we can write the ground state and the corresponding energy of our system,
\begin{eqnarray}
| G_{W} \rangle &=& \hat{T} |0\rangle \bigotimes_{j=1}^{N}  | g \rangle_{j} \nonumber \approx \frac{| 0 \rangle - \eta |2\rangle}{\sqrt{\eta^2 + 1}} \bigotimes_{j=1}^{N}  | g \rangle_{j} ,  \\
E_{G_{W}}  &=& \hbar \left( \tilde{\kappa} - \frac{N \omega_{a}}{2} \right),
\end{eqnarray}
where $|g\rangle_j$ denotes the ground state for the $j$-th TLAs. The ground state is a pure separable state and independent on the size of the atomic ensemble. Unlike the Dicke model, here the field is in a superposition of vacuum $|0\rangle$ and two-photon $|2\rangle$ states due to the degenerate parametric process.

\section{Strong coupling regime, $\lambda \gg \omega_{a}$.}

In the weak coupling regime, the ground state is well described by a finite superposition of Fock states times the two-level system ground state; while in the strong coupling regime, it is possible to consider the field in a coherent state and try to find the corresponding ensemble state. By substituting the photon creation and annihilation operators by their expectation values, the  Hamiltonian in Eq.\eqref{eq:DmH}  becomes
\begin{eqnarray} \label{eq:SemiClassH}
\hat{H}_{S}&=& \hbar \omega_{f} |\alpha|^2  + \hbar \kappa  \alpha_{R}^2 +  \hbar \omega_{a} \hat{J}_{z} + \frac{2 \hbar \lambda}{\sqrt{N}} \alpha_{R} \hat{J}_{x},
\end{eqnarray}
where the complex coherent state parameter is defined as $\alpha = \alpha_{R} + \imath \alpha_{I}$. It is possible to arrange this semi-classical Hamiltonian in Eq.\eqref{eq:SemiClassH} as a nested array of tensor products of the form 
\begin{eqnarray}
\hat{H}_{S} = \hbar \left( \omega_{f} |\alpha|^2   \kappa \alpha_{R}^2 \right) \mathbb{I}_{2^{N}} 
+ \left\{ \left[ \left( \ldots \right) \otimes \mathbb{I}_{2} + \mathbb{I}_{2^{N-2}} \otimes \hat{H}_{2} \right] \otimes \mathbb{I}_{2} + \mathbb{I}_{2^{N-1}} \otimes \hat{H}_{2} \right\},  
\end{eqnarray}
where the symbol $\mathbb{I}_{d}$ represents the unit matrix of dimension $d$ and the auxiliary matrix of dimension two is
\begin{eqnarray}
\hat{H}_{2} = \hbar \left( \frac{\omega_{a}}{2} \hat{\sigma}_{z}  + \frac{2 \lambda \alpha_{R}}{\sqrt{N}}  \hat{\sigma}_{x} \right).
\end{eqnarray}
Thus, the ground state energy is found
\begin{eqnarray} \label{eq:SemiClassEnG}
E_{G_{S}} &=&\hbar \left[ \omega_{f} |\alpha|^2 +  \kappa  \alpha_{R}^2 - \frac{N}{2} \sqrt{\omega_{a}^2 + 16 \lambda^{2} \alpha_{R}^2 / N}\right].\nonumber\\
\end{eqnarray}
In order to calculate the critical coupling value, we optimize this ground state energy for the real and imaginary parts of the coherent state parameter, $\alpha$, and find the following self-consistency equations,
\begin{eqnarray} \label{eq:ReImAlpha}
\alpha_{R}^2 &=& \frac{ N \left[ 16 \lambda^{4} - \omega_{a}^2 (\omega_{f} + 4 \kappa)^2 \right]}{4 \lambda^2 (\omega_{f} + 4 \kappa)^2}, \nonumber \\
\alpha_{I} &=& 0.
\end{eqnarray}
The phase transition in the strong coupling regime occurs at the critical value given by the expression,
\begin{eqnarray} \label{eq:SCritC}
\lambda_{S} = \frac{1}{2} \sqrt{\omega_{a} (\omega_{f} + 4 \kappa)}.
\end{eqnarray}
Although a finite size has been assumed for the atomic ensemble, this critical coupling value found in the strong coupling regime, Eq.\eqref{eq:SCritC}, is in accord with that derived from the free energy by using the thermodynamic limit method for an infinitely large two-level system ensemble \cite{Hepp1973p2517,Wang1973p831, Rzazewski1975p432} for the reason that in both cases the field is assumed to be in a coherent state.

The mean-field constrain set, Eq.\eqref{eq:ReImAlpha}, approximates, in the strong coupling regime, $\lambda \gg \lambda_{S}$, the following ground state and ground state energy,
\begin{eqnarray} 
|G_{S} \rangle &=& |\alpha \rangle \bigotimes_{j=1}^{N}  | v \rangle_{j}, \nonumber \\
E_{G_{S}} &=& - \frac{\hbar N \left[ 16 \lambda^{4} + \omega_{a}^2 (\omega_{f} + 4 \kappa)^2 \right]}{16 \lambda^{2} \left( \omega_{f} + 4 \kappa \right)},
\label{eq:GSEPT} 
\end{eqnarray}
where the auxiliary two-level state is defined as
\begin{eqnarray}
| v \rangle &=& \frac{1}{\sqrt{\beta^2 + 1}} \left( | g \rangle + \beta | e \rangle  \right), \nonumber  \\
\beta &=& \frac{\omega_{a} \left( \omega_{f} + 4 \kappa \right) - 4 \lambda^{2}}{\sqrt{16 \lambda^{4} - \omega_{a}^2 \left( \omega_{f} + 4 \kappa \right)^{2} }},.
\label{eq:aux}
\end{eqnarray}
Again, as expected, the ground state is a pure separable state; here the difference is that each component of the ensemble is in a superposition of the ground, $|g\rangle$, and excited states, $|e\rangle$. Furthermore, for a coupling parameter larger than the nonlinear parameter, $\lambda \gg \kappa$, the auxiliary state is the balanced superposition $|v\rangle = (|g\rangle - |e\rangle)/\sqrt{2}$ with null population difference, $\langle \hat{\sigma}_{z} \rangle =0$.

\section{Moderate coupling regime}

Besides the weak and strong coupling regimes, where the ground states are both separable states, we apply a direct numerical calculation to find the ground state  in the moderate coupling regime. In the simulation, each and every single two-level system is taken to be indistinguishable from each other and the angular momentum eigenstates basis is used, 
\begin{eqnarray}
\hat{J}_{z} | N/2, m \rangle   &=&  m    | N/2, m \rangle, \nonumber \\
\hat{J}_{\pm} | N/2, m \rangle &=&  \sqrt{ \frac{N}{2} \left(\frac{N}{2} + 1 \right) - m (m\pm 1)}    | N/2, m \pm 1 \rangle , \nonumber\\ 
\end{eqnarray}
where the Dicke state $|N/2, m\rangle$ is the superposition of all possible ensemble states with $N/2 + m$ two-level systems in the excited state and the rest, $N/2 - m$, in the ground state, such that $m = -N/2, -N/2 + 1, \ldots, N/2-1, N/2$. 

As the eigenstate of a truncated version of the studied Hamiltonian, Eq.\eqref{eq:DmH}, can be easily verified to be, or not, an eigenstate for the exact full Hamiltonian, the numerical approach taken here consists on assessing a maximum number of allowed excitations for the field, $n$, set to deliver at most a maximum error parameter, $\epsilon = |E_{1} - \langle \hat{H} \rangle_j|/|\langle \hat{H} \rangle_j|$, for a wide range of the phase space set by the coupling and nonlinear parameters, $(\lambda,\kappa)$ in that order. The set $\{ (E_{j}, |\psi_{j} \rangle )\}$ are the numerical eigenvalues and eigenstates, respectively, of the truncated Hamiltonian sorted in ascending order, $E_{j} \le E_{j+1}$ for $j = 1, \ldots, (N+1)(n+1)$, and the notation $ \langle \cdot \rangle_j \equiv \langle \psi_{j}|\cdot|\psi_{j} \rangle$ is used.  In addition, a degeneracy parameter  $\epsilon_{d} = |E_{n} - E_{1}| / |E_{1}|$ is established to discriminate between non-degenerate and degenerate ground states. In the latter case, the proper ground state is constructed as the normalized direct sum of the degenerate eigenstates. 

In the following numerical analysis, the error and degeneracy parameter are set to the values $\epsilon \le 10^{-10}$ and $\epsilon_{d} \le 10^{-10}$. A maximum of two hundred excitations for the field, $n = 200$, is set in accordance.
Numerical results for the on-resonance, $\omega_{f} = \omega_{a}$, and off-resonance, $\omega_{f} \in [0.85,1) \omega_{a}$, case  are performed for an assorted collection of parameters, $N \in [2,6]$, $\lambda/\sqrt{N} \in [0,5] \omega_{a}$, $\kappa \in [0,5] \omega_{a}$.
For the sake of brevity, only those results pertaining a bipartite and pentapartite ensemble are shown in Fig. \ref{fig:Figure1} for $N=2$ atoms and in Fig. \ref{fig:Figure2} for $N = 5$ atoms, respectively.  

The mean value of the z-component of the angular momentum,  $\langle \hat{J}_{z} \rangle$, which will be called population difference from now on, is shown in Fig. \ref{fig:Figure1}(a).
Simulation results reveal that, as derived in the weak coupling regime, $\lambda \ll \omega_{a}$, the population difference is minimal, $\langle \hat{J}_{z} \rangle = - N/2$, \textit{i.e.},   each and every two-level system is in its ground state, and independent of the nonlinear parameter $\kappa$. 
Also, for a sufficiently large coupling, $\lambda \gg \omega_{a}$, along with an adequate nonlinear parameter $\kappa \ll (4 \lambda^2 - \omega_{a} \omega_{f})/ 4\omega_{a}$ such that $\lambda \gg \kappa$,  the population difference is null, $\langle \hat{J}_{z} \rangle = 0$, which relates to the ground state derived in the strong coupling regime, Eq. (\ref{eq:aux}), under the aforementioned restrictions.
Fig. \ref{fig:Figure1}(b) shows that the numerical mean photon number for the field, $\langle \hat{n} \rangle$, is in agreement with the general behavior found in the analytical results; \textit{i.e.}, the field is in the vacuum field state, with a small two-photon component depending on the strength of the nonlinear parameter, for the weak coupling regime and in a coherent state, with mean photon number $|\alpha|^2$, for the strong coupling regime.

In order to demonstrate the existence of entanglement for the studied Hamiltonian in a moderate coupling regime, we calculate the maximum shared bipartite concurrence following the entangled web approach \cite{Koashi2000p050302(R)}, Fig. \ref{fig:Figure1}(c), and the field-ensemble entanglement probed through von Neumann entropy of the reduced two-level ensemble, also known as entropy of entanglement \cite{Nielsen2000}, Fig. \ref{fig:Figure1}(d).
Non-zero regions for  both the bipartite concurrence and the entropy of entanglement are found in between the separable states corresponding to the weak and strong coupling regimes, approximately delimited by the black lines in the Fig. \ref{fig:Figure1}(c) and (d).
It is possible to see that the maximum shared bipartite concurrence locates in the upper diagonal region, Fig. \ref{fig:Figure1}(c), indicating that the entanglement shared between the ensemble components occurs due to an approximately equal balance between the linear atom-photon and nonlinear photon-photon interactions.
Instead, the entropy, which has its maximum value below the diagonal region, Fig. \ref{fig:Figure1}(d), shows a maximum entanglement between the two-level system ensemble and the photon field due to a larger atom-photon interaction strength.

A shortcoming of the numerical approach shows up at this point.
The area of zero entropy below the entangled phase is inversely proportional to the value of the degeneracy parameter, $\epsilon_{d}$, mentioned above.
Also, the error parameter, $\epsilon$, increases as the nonlinear parameter $\kappa$ goes to zero. These shortcomings appear due to the truncation of the Hilbert space for solving the eigenvalue problem.
When the counter-rotating and diamagnetic like terms are neglected, the system is confined to certain finite subspaces and the numerical approach does not present this problems \cite{Buzek2005p163601}.
Numerical results might be improved by allowing a larger maximum for the maximum excitation of the field, optimizing the code, or effecting a customized analytical progressive diagonalization scheme based on those presented in references \cite{Pan2009p044306,Pan2010}.\\
Nevertheless, the current approach allows the calculation of the states for the field and atomic ensemble up to the desired precision.
In the second row of Fig. \ref{fig:Figure1}, we show the photon number probability distributions, $P(n)= |\langle n | \psi_{g} \rangle|^2$, related to the four markers, labeled from $A$ to $D$, along a constant coupling parameter, $\lambda=3.323$, represented by the solid line in Fig. \ref{fig:Figure1}(c).
In the absence of nonlinear coupling, $\kappa = 0$, a Poissonian photon number distribution is discovered in Fig. \ref{fig:Figure1}(e). By calculating the field quadratures variances $\langle\Delta \hat{X}^2\rangle$ and $\langle\Delta \hat{Y}^2\rangle$, with the field quadratures defined as $\hat{X}=\hat{a}^{\dagger}+\hat{a}$ and $\hat{Y}=\imath (\hat{a}^{\dagger} - \hat{a})$, the field squeezing is probed. In this case, $\kappa = 0$, the field is in a coherent state, as expected; \textit{i.e.}, the field quadratures mean values and their uncertainty relation have all a value of one.
With a small value of the nonlinear coupling strength,  $\kappa = 0.3$, the statistics for the photon number distribution becomes sub-Poissonian, shown in Fig. \ref{fig:Figure1}(f).
The field is in a squeezed coherent state as the uncertainty relation for the field quadratures remains minimal but the variance $\langle\Delta \hat{X}^2\rangle$ increases as $\langle\Delta \hat{Y}^2\rangle$ decreases.
By increasing the nonlinear coupling,  $\kappa = 2.4$, an oscillating photon number distribution is found in Fig. \ref{fig:Figure1}(g).
Now, the quadrature squeezing seems to be reversed and the variance $\langle\Delta \hat{Y}^2\rangle$ is smaller than $\langle\Delta \hat{X}^2\rangle$ and close to a value of one.
Also, as the value for the quadratures uncertainty relation is more than one, the field is no longer in a coherent state.
For further increasing of the nonlinear coupling, $\kappa =4.8$, the oscillating photon number distribution remains, Fig. \ref{fig:Figure1}(h),
the variance $\langle\Delta \hat{Y}^2\rangle$ is further squeezed, and the field is not a coherent state but shows a tendency to become the superposition of the vacuum and two-photon state.

Besides the photon number probability distributions, the corresponding z-component angular momentum probability distributions,  $P(m)= |\langle m | \psi_{g} \rangle|^2$, are shown in the third row of Fig.\ref{fig:Figure1}, in the same order related to the four points $A$ to $D$ along the solid line in Fig. \ref{fig:Figure1} (c).
We calculate the mean values and variances for the three momentum operators, $\langle\hat{J}_{i}\rangle$ and $\langle\Delta \hat {J}_{i}^2\rangle$ for $i = x,y,z$, as well as the uncertainty relation between the population difference and the dipole phase, $4 \langle\Delta \hat {J}_{z}^2\rangle \langle\Delta \hat {\Phi}^2\rangle \ge 1$ where $\langle\Delta \hat {\Phi}^2\rangle = \langle\Delta \hat {J}_{y}^2\rangle / \langle \hat {J}_{x} \rangle^2$.
Again, by increasing the nonlinear coupling strength, Fig. \ref{fig:Figure1} (i-l), the atomic state changes from a coherent atomic state in the absence of the nonlinear parameter, to a squeezed coherent atomic state for a small nonlinear parameter.
For a larger nonlinear coupling strength, the squeezed atomic states becomes a state where the minimal Dicke state, $|N, -N/2\rangle$, predominates.  
Our simulation results indicate the co-existence of squeezed fields and squeezed atomic ensembles in the moderate coupling regime.
The field and atomic statistics for the points discussed above, approximated to three decimals for the sake of space, are shown in Table \ref{tab:Table1}.

As the number of two level systems increases, \textit{e.g.}, $N = 5$ in Fig. \ref{fig:Figure2}, the maximum bipartite entanglement shared between ensemble components seems to be inversely proportional to the ensemble size and the region of entanglement decreases.
In the second and third rows of Fig. \ref{fig:Figure2}, similar photon and atomic statistics from Poissonian, sub-Poissonian, to oscillating photon number distributions for the field and from the coherent to squeezed atomic ensembles, respectively, are demonstrated along a constant coupling parameter, $\lambda=3.019$.

\section{Conclusion}
Two phase transitions for the ground state were found for a finite size Dicke Hamiltonian plus counter-rotating and quadratic field terms, corresponding to the weak and strong coupling regimes.
The ground states before and after these transitions are analytically found to be pure separable states, thus there exists no entanglement in the system, identified from each other by both the state of the field and two-level system ensemble; \textit{i.~e.},  the superposition of the vacuum and two photon field states times all the components of the ensemble in the ground state, for couplings lesser than the weak critical coupling, and a non-vacuum coherent field state times all the components of the ensemble in a superposition of ground and excited states, for couplings larger than the strong critical coupling.

In between these extremes, the ground state presents both ensemble--field entanglement and bipartite entanglement between the ensemble components. Results on ensemble bipartite entanglement behave as expected, the degree of maximum shared pairwise entanglement decreases as the number of entangled pairs in the two-level ensemble increases; \textit{i.e.}, for a sufficiently large ensemble, \textit{e.~g.}, the infinitely large ensemble considered in the thermodynamic limit, the maximum shared bipartite entanglement will tend to zero and there will be no intermediate region between the weak and strong regimes. Thus, the phase space region for which the ground state of the system is entangled is directly related to the finite size of the system.

\begin{acknowledgments}
This work was supported by the National Tsing-Hua Univesity under contract No. 98N2309E1.
\end{acknowledgments}

 \newpage

\begin{widetext}
\begin{center}
\begin{figure}[ht]
\includegraphics[width = 6in]{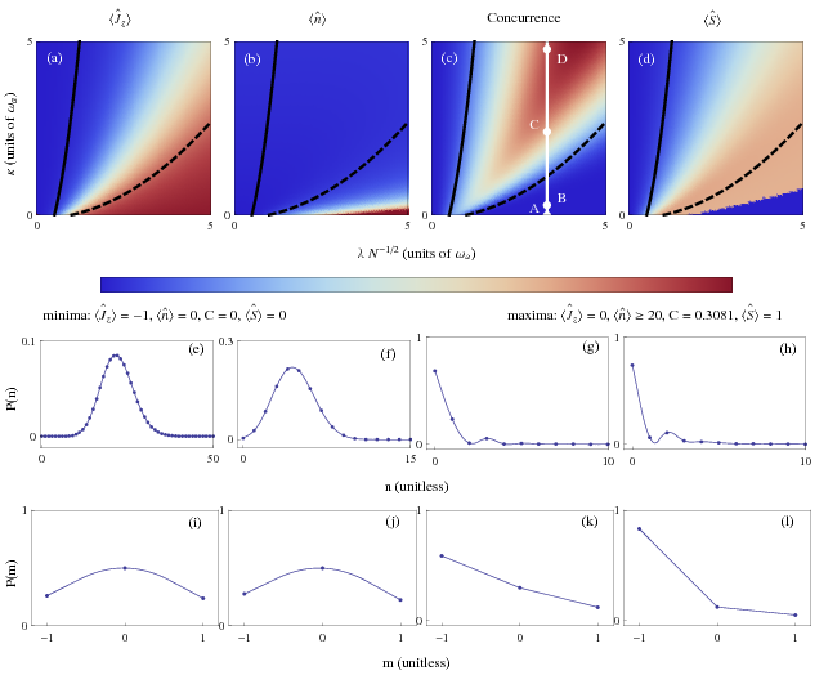}
\caption{(Color online) The phase diagram of our finite-size Dicke Hamiltonian in the parameter space of the linear photon-atom coupling strength, $\lambda$, and the nonlinear photon-photon interaction strength, $\kappa$. (a) The mean value for the atomic z-component, $\langle \hat{J}_z \rangle$, (b) the average photon number for the field, $\langle \hat{n} \rangle$, (c): the bipartite concurrence, and (d) the entropy of entanglement, $\langle \hat{S}\rangle$, are calculated for the case of $N = 2$. The corresponding minima and maxima values for the color legend are shown below. The field photon number and atomic angular momentum probability distributions along the solid line in (c) are shown in (e-h) and (i-l), ordered according to the markers $A$-$D$, respectively.} \label{fig:Figure1}
\end{figure}
\end{center}
\end{widetext}

\begin{widetext}
\begin{center}
\begin{figure}[ht]
\includegraphics[width = 6in]{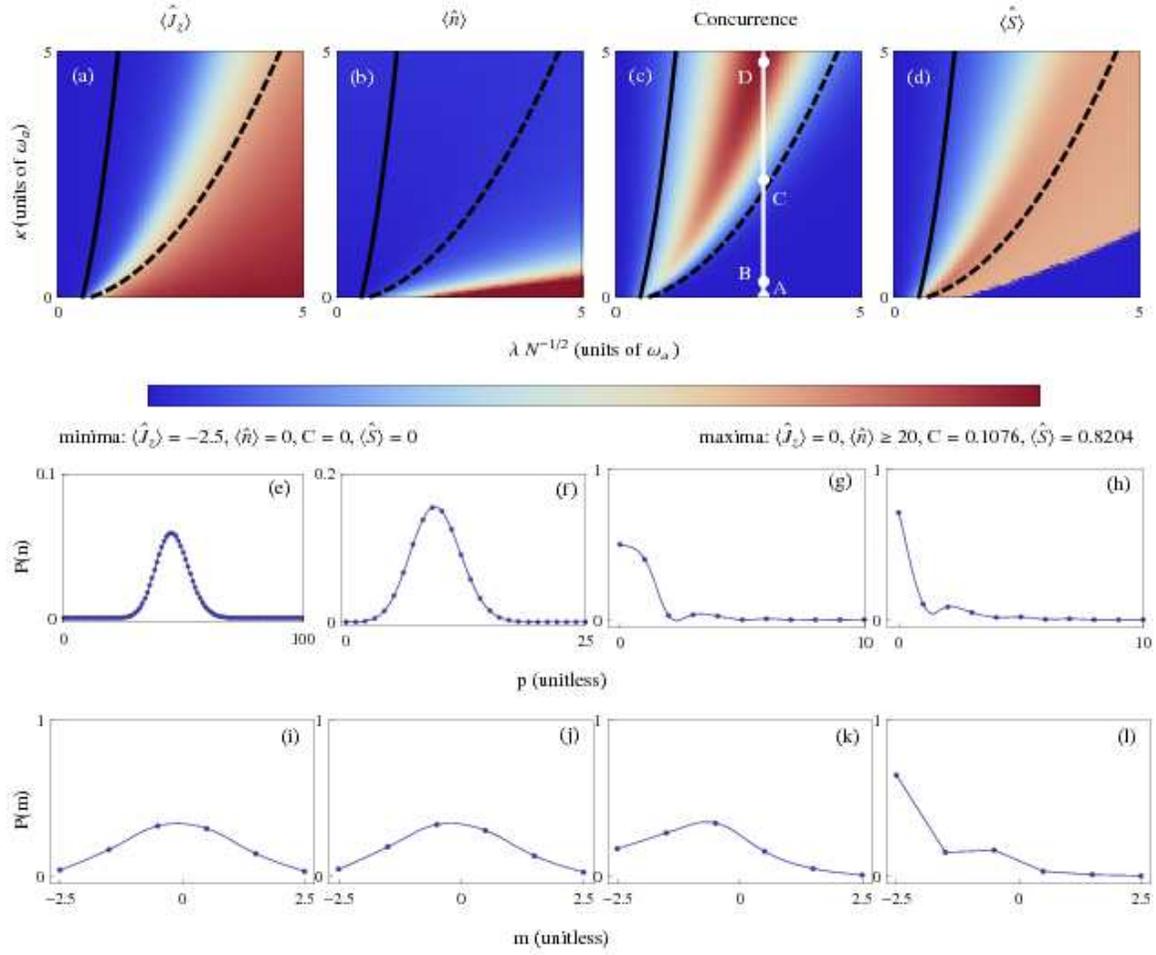}
\caption{(Color online) Same as Fig.\eqref{fig:Figure1}, but for the case of $N = 5$.} \label{fig:Figure2}
\end{figure}
\end{center}
\end{widetext}

\begin{widetext}
\begin{center}
\begin{table}
\begin{tabular}{|c|c|c|c|c|c|c|c|c|}
\hline
&\multicolumn{4}{c|}{ 2 TLS, Fig. \ref{fig:Figure1}(e-l) }&\multicolumn{4}{c|}{ 5 TLS, Fig. \ref{fig:Figure2}(e-l)} \\ 
&\multicolumn{4}{c|}{ $\lambda = 3.323 $ }&\multicolumn{4}{c|}{ $\lambda =  3.019 $} \\ \hline 
& A & B & C & D & A & B & C & D \\ 
& $\kappa = 0 $ & $\kappa = 0.3$ & $\kappa = 2.4$ & $\kappa = 4.8$ & $\kappa = 0$ & $\kappa = 0.3$ & $\kappa = 2.4$ & $\kappa = 4.8$ \\ \hline 
$\langle \hat{n} \rangle$ & 22.078 & 4.590 & 0.502 & 0.678 & 45.528 & 9.417 & 0.731 & 0.707\\ \hline 
$\langle \Delta \hat{n}^2 \rangle$ & 22.084 & 3.155 & 1.011 & 2.135 & 45.545 & 6.416 & 1.153 & 2.162 \\ \hline 
$\langle \Delta \hat{X}^2 \rangle$& 1.000 & 0.676 & 1.824 & 2.941 & 1.000 & 0.676 & 1.645 & 2.609\\ \hline 
$\langle \Delta \hat{Y}^2 \rangle$& 1.000 & 1.481 & 0.869 & 0.355 & 1.000 & 1.480 & 1.087 & 0.426\\ \hline 
$\langle \Delta \hat{X}^2 \rangle \langle \Delta \hat{Y}^2 \rangle$& 1.000 & 1.000 & 1.586 & 1.045 & 1.000 & 1.000 & 1.789 & 1.112 \\ \hline 
$\langle \hat{J}_{x} \rangle$&-1.000&0.000&0.000&0.000&-2.499&2.495&0.000&0.000 \\ \hline 
$\langle \Delta \hat{J}_{x}^2 \rangle$&0.000&-0.052&0.903&0.756&0.001&0.005&5.630&3.576 \\ \hline 
$\langle \hat{J}_{y} \rangle$&0.000&0.000&0.000&0.000&0.000&0.000&0.000&0.000\\ \hline 
$\langle \Delta \hat{J}_{y}^2 \rangle$&0.5000&0.500&0.392&0.364&1.250&1.250&1.147&0.719 \\ \hline 
$\langle \hat{J}_{z} \rangle$& -0.023&-0.512&-0.466&-0.781&-0.068&-0.153&-0.861&-1.902\\ \hline 
$\langle \Delta \hat{J}_{z}^2 \rangle$&0.500&0.5000&0.487&0.269&1.250&1.246&1.232&0.837 \\ \hline 
$ 4 \langle \Delta \hat{J}_{z}^2 \rangle \langle \Delta \hat{\Phi}^2 \rangle $&1.000&1.151$\times 10^{47}$&7.515$\times 10^{54}$&5.887$\times 10^{51}$&1.000&1.000&7.180$\times 10^{38}$&1.682$\times 10^{41}$ \\ \hline 
\end{tabular}
\caption{The field and atomic statistics for the markers $A$ to $D$ in Fig. \ref{fig:Figure1}(c) for $N = 2$ and in Fig. \ref{fig:Figure2}(c) for $N = 5$ two-level systems. See the text for more details.} \label{tab:Table1}
\end{table}
\end{center}
\end{widetext}

\end{document}